\documentclass[11pt, onecolumn]{IEEEtran} 
\def\BibTeX{{\rm B\kern-.05em{\sc i\kern-.025em b}\kern-.08em
     T\kern-.1667em\lower.7ex\hbox{E}\kern-.125emX}}

\usepackage{epsf,cite,amsmath,amscd,amssymb,graphicx,latexsym,multicol}

\newtheorem{theorem}{Theorem}[section]

\newtheorem{definition}[theorem]{Definition}

\centerfigcaptionstrue

\begin{document}
\bibliographystyle{ieeetr}

\title{{Energy-Efficient Resource Allocation in Wireless Networks: An overview of game-theoretic approaches\footnote{This research was supported by the National
Science Foundation under Grant ANI-03-38807.}}\vspace{.5cm}}
\author{Farhad Meshkati, H. Vincent Poor, and Stuart C. Schwartz\vspace{0.5cm}\footnote{F. Meshkati was with the Department of
Electrical Engineering at Princeton University. He is currently
with QUALCOMM Inc., 5775 Morehouse Dr., San Diego, CA 92121 USA
(e-mail: meshkati@qualcomm.com).} \footnote{H. V. Poor and S. C.
Schwartz are with the Department of Electrical Engineering,
Princeton University, Princeton, NJ 08544 USA (e-mail:
{\{poor,stuart\}@princeton.edu}).}}

\maketitle

\begin{abstract}
An overview of game-theoretic approaches to energy-efficient
resource allocation in wireless networks is presented. Focusing on
multiple-access networks, it is demonstrated that game theory can
be used as an effective tool to study resource allocation in
wireless networks with quality-of-service (QoS) constraints. A
family of non-cooperative (distributed) games is presented in
which each user seeks to choose a strategy that maximizes its own
utility while satisfying its QoS requirements. The utility
function considered here measures the number of reliable bits that
are transmitted per joule of energy consumed and, hence, is
particulary suitable for energy-constrained networks. The actions
available to each user in trying to maximize its own utility are
at least the choice of the transmit power and, depending on the
situation, the user may also be able to choose its transmission
rate, modulation, packet size, multiuser receiver, multi-antenna
processing algorithm, or carrier allocation strategy. The
best-response strategy and Nash equilibrium for each game is
presented. Using this game-theoretic framework, the effects of
power control, rate control, modulation, temporal and spatial
signal processing, carrier allocation strategy and delay QoS
constraints on energy efficiency and network capacity are
quantified.
\end{abstract}


\section{Introduction and Motivation}

Future wireless networks are expected to support a variety of
services with diverse quality-of-service (QoS) requirements. For
example, a mixture of delay-sensitive applications (e.g., voice
and video teleconferencing) and delay-tolerant ones (e.g., web
browsing and file downloading) must be supported. Given that the
two principal wireless network resources, i.e., \emph{bandwidth}
and \emph{energy}, are scarce, the main challenge in designing
wireless networks is to use network resources as efficiently as
possible while providing the QoS required by the users.

Game-theoretic approaches to radio resource allocation have
recently attracted much attention and will be the focus of this
article. We will show that game theory can be used as a unifying
framework to study radio resource management in a variety of
wireless networks with different service criteria. Our focus will
be on infrastructure networks where users transmit to a common
concentration point such as a base station in a cellular network
or an access point. Since most of the terminals in a wireless
network are battery-powered, energy efficiency is crucial to
prolonging the life of the terminals. Also, in most practical
scenarios, distributed algorithms are preferred over centralized
ones. Centralized algorithms tend to be complex and not easily
scalable. Therefore, throughout this article, we focus on
distributed algorithms with emphasis on energy efficiency. Using a
game-theoretic framework, we demonstrate the impact of advanced
signal processing on energy efficiency and network capacity. The
tradeoffs among throughput, delay, network capacity and energy
efficiency are also discussed. The ideas presented in this paper
can also be applied to wireless ad hoc networks, however, the
topic is beyond the scope of this article (see \cite{Srivastava05}
for applications of game theory to ad hoc networks).

It should be noted that, recently, tools from optimization theory
have also been employed to study resource allocation in wireless
networks using the network utility maximization framework proposed
in \cite{Kelly98} (see for example \cite{CLC06}). While there is
considerable overlap between the game-theoretic and
optimization-theoretic approaches, game theory tends to focus on
the multiuser competitive nature of the problem and on the users'
interaction.

The rest of this article is organized as follows. In
Section~\ref{gametheory}, we describe how game theory can be used
for studying radio resource management in wireless networks. The
choice of the utility function is discussed in
Section~\ref{utilityfunction}. In Section~\ref{PCG}, we present a
family of power control games for energy-efficient resource
allocation in wireless CDMA networks. Finally, discussions and
conclusions are given in Section~\ref{conclusions}.

\section{Game Theory for Radio Resource Management} \label{gametheory}

Game theory is a mathematical tool for analyzing the interaction
of two or more decision makers. Game theory has been used in a
variety of fields such as economics, political science, and
biology \cite{StraffinBook}. A (strategic) game consists of three
components: a set of players, the strategy set for each player and
a utility (payoff) function for each player measuring the degree
of ``happiness" of the player \cite{FudenbergTiroleBook91}.
Recently, game theory has also been used in telecommunications and
particularly wireless communications (see for example \cite{AB02,
MackenzieWicker01, ORS93, Altman06}). The users' interaction in a
wireless network can be modeled as a game in which the users'
terminals are the players in the game competing for network
resources (i.e., bandwidth and energy). Any action taken by a user
affects the performance of other users in the network. Game theory
is the natural tool for studying this interaction.

Since our focus in this article is on distributed schemes, we will
concentrate on non-cooperative games. Let
${\mathcal{G}}=\left[\mathcal{K}, \{\mathcal{A}_k\}, u_k\right]$
represent a game where $\mathcal{K}=\{1,\cdots,K\}$ is the set of
players/users, $\mathcal{A}_k$ is the set of actions (strategies)
available to user $k$, and $u_k$ is the utility (payoff) function
for user $k$. In a non-cooperative game, each user seeks to choose
its strategy in such a way as to maximize its own utility, i.e.,
\begin{equation}\label{eq1}
    \max_{a_k\in \mathcal{A}_k} u_k  \ \ \textrm{for} \ \
    k=1,\cdots, K .
\end{equation}
For such a game, we first need to define two important concepts,
namely, a \emph{Nash equilibrium }and \emph{Pareto optimality}
\cite{FudenbergTiroleBook91}.

\begin{definition}
A Nash equilibrium (NE) is a set of strategies, $(a_1^*,\cdots,
a_K^*)$, such that no user can unilaterally improve its own
utility, that is,
\begin{equation}\label{eq2}
    u_k(a_k^*, a_{-k}^*)\geq u_k(a_k,a_{-k}^*) \ \ \textrm{for
    all} \ a_k\in \mathcal{A}_k \ \textrm{and} \ \ k=1,\cdots,K ,
\end{equation}
where $a_{-k}^*=(a_1^*,\cdots, a_{k-1}^*, a_{k+1}^*,\cdots,
a_K^*)$.
\end{definition}

A Nash equilibrium is a \emph{stable} outcome of ${\mathcal{G}}$.
At NE, no user has any incentive to change its strategy.

\begin{definition}
A set of strategies, $(\tilde{a}_1,\cdots, \tilde{a}_K)$ is
Pareto-optimal if there exists no other set of strategies for
which one or more users can improve their utilities without
reducing the utilities of other users
\end{definition}

It should be noted that our focus throughout this paper is on pure
strategies. However, one could also allow for the users to have
mixed strategies. In such a case, each user assigns a probability
distribution to its pure strategies and then it chooses a pure
strategy based on the probability distribution. A non-cooperative
game may have no pure-strategy Nash equilibrium, one equilibrium
or multiple equilibria. Also, in many cases, a NE may not be
Pareto-efficient (Pareto-optimal).

\begin{figure}[t]
\centering
\includegraphics[width=2in]{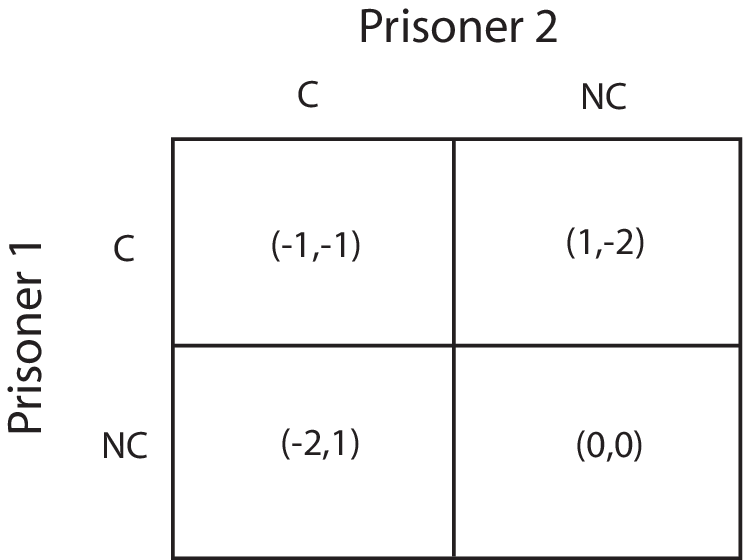}
\caption{The matrix-form representation of the Prisoner's Dilemma
game.} \label{fig1}
\end{figure}

As an example, consider the following two-player game called the
\emph{Prisoner's Dilemma} \cite{StraffinBook}. The two players are
two prisoners that have been arrested for a joint crime. They are
taken into separate rooms and are given the options to either
\emph{confess} ($C$) to the crime or \emph{not confess} ($NC$).
Each prisoner is told that if they both confess, each gets a light
sentence (i.e., payoff of $-1$). If neither confesses, both will
go free (i.e., payoff of $0$). If one of them confesses and the
other one does not, the confessor will get a reward (i.e., payoff
of $+1$) and the other prisoner will get a heavy sentence (i.e.,
payoff of $-2$). The actions and the corresponding payoffs of the
players are shown in Fig.~\ref{fig1}. Since the two prisoners are
in separate rooms and hence are not able to cooperate, the
payoff-maximizing selfish strategy for each of them is to confess.
It can easily be verified that $(C,C)$ is the unique Nash
equilibrium of this game. Furthermore, this equilibrium is not
Pareto-efficient since choosing $(NC,NC)$ would result in a larger
payoff for both players. However, this would require cooperation
between the two prisoners. Hence, it is evident from this example
that there is a clear conflict between individual rationality and
social welfare.

In this article, we provide an overview of game-theoretic
approaches to energy-efficient resource allocation in wireless
networks. Consider the uplink of a direct-sequence code-division
multiple-access (DS-CDMA) network where each user wishes to
locally and selfishly choose its action in such a way as to
maximize its own utility while satisfying its QoS requirements.
Depending on the situation, the actions open to each user in
trying to maximize its own utility can be, for example, the choice
of its transmit power, transmission rate, modulation, packet size,
multiuser receiver, multi-antenna processing algorithm, or carrier
allocation strategy. The strategy chosen by a user affects the
performance of other users in the network through multiple-access
interference. There are several important questions to ask
regarding game $\mathcal{G}$. First of all, what is a reasonable
choice of utility function? Secondly, given the utility function,
what strategy must a user choose in order to maximize its own
utility (i.e., best-response strategy)? If every user in the
network selfishly and locally picks its best-response strategy,
will there be a steady-state solution where no user can
unilaterally improve its utility (i.e., Nash equilibrium)? If such
a steady-state solution exits, is it unique? How does the
performance of such a non-cooperative approach compare with a
cooperative scheme?

Let us consider the uplink of a synchronous DS-CDMA network with
$K$ users.\footnote{For the sake of simplicity, it is common to
focus on a synchronous CDMA system. Many of the results presented
in this paper can be generalized to asynchronous systems as well.}
Assuming quasi-static fading, the signal received by the uplink
receiver (after chip-matched filtering) sampled at the chip rate
over one symbol duration can be expressed as
\begin{equation}\label{eq3}
    {\mathbf{r}} = \sum_{k=1}^{K} \sqrt{p_k h_k} \ b_k {\mathbf{s}}_k +
    {\mathbf{w}} ,
\end{equation}
where $p_k$, $h_k$, $b_k$ and ${\mathbf{s}}_k$ are the transmit
power, channel gain, transmitted bit and spreading sequence of
user $k$, respectively, and $\mathbf{w}$ is the noise vector which
include other-cell interference and is assumed to be Gaussian with
mean $\mathbf{0}$ and covariance $\sigma^2 \mathbf{I}$. Throughout
this article, we study distributed resource allocation in such a
wireless network by presenting several (non-cooperative) power
control games in which users choose their strategies in such a way
as to maximize their utilities. The emphasis will be mainly on
energy efficiency. It should be noted that in the power control
games under consideration, the actions available to the users are
not limited to the choice of transmit power. Depending on the
situation, the users may also choose their transmission rates,
modulation schemes, packet sizes, multiuser receivers,
multi-antenna processing algorithms, or carrier allocation
strategies. Furthermore, \emph{cross-layer} resource allocation
can be achieved by expanding the strategy sets of the users over
multiple layers in the OSI protocol stack or by defining the
users' utility functions such that performance measures across
multiple layers are included.

\section{Utility Function} \label{utilityfunction}

Based on the discussions in the previous section, the choice of
the utility function has a great impact on the nature of the game
and how the users choose their actions. For resource allocation in
wireless data networks, several different utility functions have
been used in the literature.

When maximizing the spectral efficiency is the main goal, it is
common to define the user's utility as a logarithmic, concave
function of the user's signal-to-interference-plus-noise ratio
(SIR)\cite{Alpcan, Gunturi03}, i.e.,
\begin{equation}\label{eq4}
    u_k = \zeta_k \log (1+\gamma_k) ,
\end{equation}
where $\gamma_k$ is the SIR for user $k$, and $\zeta_k$ is a
constant which is in general user-dependent. This utility function
is proportional to the Shannon capacity for the user treating all
interference as white Gaussian noise.  In addition, a pricing
function is introduced to prevent the users from always
transmitting at full power. In many cases, the pricing function is
assumed to be linear in the user's transmit power. Hence, the net
utility for user $k$ is given by
\begin{equation}\label{eq5}
    \tilde{u}_k =\zeta_k \log (1+\gamma_k) - c_k p_k
\end{equation}
where $c_k$ is the pricing factor for user $k$.

The authors in \cite{Xiao01} define the utility function of a user
to be a sigmoidal function of the user's SIR.\footnote{An
increasing function is S-shaped if there is a point above which
the function is strictly concave, and below which the function is
strictly convex.} In this case, the net utility is defined as the
difference between the user's utility function and a (linear) cost
function, i.e.,
\begin{equation}\label{eq6}
    \tilde{u}_k = u_k - c_k p_k
\end{equation}
where, $c_k$ is again the pricing factor and $u_k$ is assumed to
be a sigmoidal function of $\gamma_k$.

In \cite{Koskie05}, the authors define a cost function (instead of
a utility function) and consider a game in which each user chooses
its transmit power to minimize its own cost. The cost function for
user $k$ is defined as
\begin{equation}\label{eq7}
    J_k = b_k p_k + c_k (\gamma_k^{tar} - \gamma_k)^2 ,
\end{equation}
where $b_k$ and $c_k$ are non-negative constants and
$\gamma_k^{tar}$ is the target SIR for user $k$. Note that this
cost function is convex and non-negative. Therefore, it has a
non-negative minimum.

\begin{figure}[t]
\centering
\includegraphics[width=3.5in]{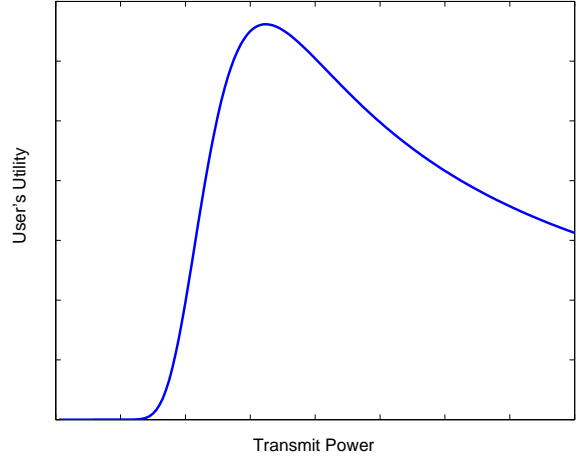}
\caption{User's utility as a function of transmit power for fixed
interference.} \label{fig2}
\end{figure}

When energy efficiency is the main concern, a good choice for the
utility function is one that measures the number of bits that can
be transmitted per joule of energy consumed. It is clear that a
higher SIR level at the output of the receiver will result in a
lower bit error rate and hence higher throughput. However,
achieving a high SIR level often requires the user terminal to
transmit at a high power which in turn results in low battery
life. This tradeoff can be captured by defining the utility
function of a user as the ratio of its throughput to its transmit
power, i.e.,
\begin{equation}\label{eq8}
    u_k = \frac{T_k}{p_k} \ .
\end{equation}
Throughput here is the net number of information bits that are
transmitted without error per unit time (this sometimes is
referred to as \emph{goodput}). It can be expressed as
\begin{equation}\label{eq9}
   T_k = R_k f(\gamma_k)  ,
\end{equation}
where $R_k$ and $\gamma_k$ are the transmission rate and the SIR
for the $k$th user, respectively; and $f(\gamma_k)$ is the
``efficiency function" which represents the packet success rate
(PSR).  The assumption here is that if a packet has one or more
bit errors, it will be retransmitted. This utility function, which
has units of \emph{bits/joule}, represents the total number of
reliable bits that are delivered to the destination per joule of
energy consumed. It captures very well the tradeoff between
throughput and battery life and is particularly suitable for
applications where energy efficiency is more important than
achieving a high throughput. The utility function in \eqref{eq8}
was introduced in \cite{Shah98, GoodmanMandayam00} and has been
used by others in scenarios in which energy efficiency is the main
concern (see for example, \cite{Saraydar01, Saraydar02,
MeshkatiTCOM05}). Obviously, $f(\gamma)$ depends on the details of
the data transmission such as modulation, coding, and packet size.
However, in most practical cases, $f(\gamma)$ is increasing and
S-shaped (sigmoidal) with $f(\infty)=1$. It is also required for
$f(\gamma)$ to be equal to zero when $\gamma=0$ to make sure that
the utility function in \eqref{eq8} does not become infinity when
$p_k=0$ (see \cite{MeshkatiTCOM05} for details). Combining
\eqref{eq8} with \eqref{eq9}, the utility function of the $k$th
user is given by
\begin{equation}\label{eq10}
    u_k = R_k \frac{f(\gamma_k)}{p_k} .
\end{equation}
Using a sigmoidal efficiency function, the shape of the utility
function in \eqref{eq10} is shown in Fig.~\ref{fig2} as a function
of the user's transmit power keeping other users' transmit powers
fixed. The utility function in \eqref{eq10} can also be used for
coded systems by modifying the efficiency function, $f(\gamma)$,
to represent the PSR for the coded system and also scaling the
transmission rate appropriately to count only the information bits
in a packet.

\section{Power Control Games} \label{PCG}

Power control is used for interference management and resource
allocation in wireless networks, especially CDMA networks. In the
uplink (from the mobile terminal to the base station), the purpose
of power control is for each user to transmit just enough power to
achieve the required QoS without causing excessive interference in
the network. Power control for CDMA systems has been studied
extensively over the past decade (see for example \cite{Zander92,
Foschini93, Yates95, HanlyTsePC, ShitzVerdu01, Leung04, CTPOJ05}).
The conventional approach has been to model power control as a
constrained optimization problem where the total transmit power is
minimized under the constraint that the users' QoS requirements
are satisfied. The QoS requirement for a user is usually expressed
as a lower bound on the user's output SIR. In \cite{Foschini93},
the authors propose a distributed algorithm for reaching the
optimum power levels. In \cite{Yates95}, a unified framework for
distributed power control in cellular networks is proposed.
Alternatively, the transmit powers of the users can be chosen in
such a way as to maximize the spectral efficiency (in bits/s/Hz).
In this approach, the optimal power control strategy is
essentially a water-filling scheme (see \cite{ShitzVerdu01}). In
\cite{CTPOJ05}, the authors use tools from geometric programming
to study power control.

Recently, game theory has been used to study power control in CDMA
systems (see, for example, \cite{JiHuang98, Shah98,
GoodmanMandayam00, Xiao01, Zhou01, Alpcan, Feng01, Sung,
Saraydar01, Saraydar02, Gunturi03, AA03}). Each user seeks to
choose its transmit power in order to maximize its utility. As
mentioned in Section~\ref{utilityfunction}, the choice of the
utility has a great impact on the nature of the game and the
resulting Nash equilibrium. In \cite{Alpcan} and \cite{Gunturi03},
the utility function in \eqref{eq5} is chosen for the users and
the corresponding Nash equilibrium solution is derived. In
\cite{Shah98} and \cite{GoodmanMandayam00}, the authors use the
utility function in \eqref{eq8} and show that the resulting Nash
equilibrium is SIR-balanced (i.e., all users have the same output
SIR). The analysis is extended in \cite{Saraydar02} by introducing
pricing to improve the efficiency of Nash equilibrium. Joint
network-centric and user-centric power control is discussed in
\cite{Feng01}. In \cite{Sung}, the utility function is assumed to
be proportional to the user's throughput and a pricing function
based on the normalized received power of the user is proposed.
S-modular power control games are studied in \cite{AA03}. In
particular, the conditions for existence and uniqueness of Nash
equilibrium for an S-modular game are discussed, and convergence
of best-response algorithms is studied.

In this section, we discuss a family of non-cooperative power
control games for resource allocation in a variety of CDMA
networks with emphasis on energy efficiency. In all these games,
the utility function measures the number of reliable bits that are
transmitted per joule of energy consumed (similar to the utility
function given in \eqref{eq10}). We discuss power control games in
which, in addition to choosing their transmit powers and depending
on the scenario, the users can choose  their uplink receivers,
MIMO processing algorithms, modulation schemes, transmission
rates, and carrier allocation strategies. We also discuss the
cases where the users seek to maximize their energy efficiency
while satisfying their delay QoS constraints.

Our focus throughout this paper is on non-cooperative
(distributed) games where each user seeks to maximize its own
utility. An alternative approach would be to maximize the sum of
the users' utilities. The solution to this problem would
correspond to a point on the Parteo-optimal frontier. However,
obtaining a closed-form solution for such an optimization problem
is usually very difficult. In addition, the solution typically
requires coordination among users and, hence, is not scalable.

\subsection{Energy-Efficient Power Control}

In \cite{Shah98} and \cite{GoodmanMandayam00}, a non-cooperative
game is proposed in which each user chooses its transmit power in
such a way as to maximize its own energy efficiency (measured in
bits/joule). To be more specific, let
${\mathcal{G}}=~[{\mathcal{K}}, \{{\mathcal{A}}_k\}, \{u_k \}]$
denote a non-cooperative game where ${\mathcal{K}}=\{1, ... , K
\}$, and $\mathcal{A}_k=[0,P_{max}]$ is the strategy set for the
$k${th} user. Here, $P_{max}$ is the maximum allowed power for
transmission. For this game, the best-response strategy for user
$k$ is given by the solution of the following maximization
problem:
\begin{equation}\label{eq11}
    \max_{p_k} u_k = \max_{p_k} R_k \frac{f(\gamma_k)}{p_k}\ \ \
    \textrm{for}  \ \  k=1,...,K  .
\end{equation}

Recall that with random spreading, the output SIR for a matched
filter receiver is given by
\begin{equation}\label{eq12}
\gamma_k = \frac{p_k h_k } {\sigma^2 + \frac{1}{N}\sum_{j\neq k}
p_j h_j } .
\end{equation}
Assuming a matched filter receiver, it is shown in
\cite{GoodmanMandayam00} that the user's utility is maximized when
the user transmits at a power level that achieves an SIR equal to
$\gamma^*$ at the output of the receiver, where $\gamma^*$ is the
unique (positive) solution of
\begin{equation}\label{eq13}
    f(\gamma) = \gamma \ f'(\gamma)  \ .
\end{equation}
It should be noted that, based on \eqref{eq13}, $\gamma^*$ depends
only on the physical-layer characteristics of the communication
such as modulation, coding and packet size. If $\gamma^*$ is not
feasible for a user, the user's utility is maximized when the user
transmits at the maximum power. Furthermore, it is shown in
\cite{Shah98} and \cite{Saraydar02} that this game has a unique
Nash equilibrium and the equilibrium is SIR-balanced, i.e, all
users have the same SIR. The existence of a Nash equilibrium is
due to the quasiconcavity of the utility as a function of the
user's transmit power.\footnote{The function $u$ defined on a
convex set $\mathcal{S}$ is quasiconcave if every superlevel set
of  $u$  is convex, i.e., $\{x \in \mathcal{S} | u(x) \geq a\}$ is
convex for every value of $a$. In other words, a function is
quasiconcave if there exists a point below which the function is
non-decreasing, and above which the function is non-increasing.}
The uniqueness of the equilibrium is because of the uniqueness of
$\gamma^*$ and the one-to-one correspondence between the users'
output SIRs and transmit powers.

The analysis is further extended in \cite{Saraydar02} to show that
this SIR-balancing NE solution is not Pareto-optimal. In
particular, it is shown that if all the users reduce their
transmit powers at the same time, the utility will improve for
every user. Based on this observation, the authors introduce a
linear pricing function and define the net utility of a user as
\begin{equation}\label{eq14}
    \tilde{u}_k = R_k \frac{f(\gamma_k)}{p_k} - c_k p_k\ \ \
    \textrm{for}  \ \  k=1,...,K  ,
\end{equation}
where $c_k$ is the pricing factor. This utility function
encourages users to transmit at a lower power level which causes
less interference for other users. A new game is proposed in which
users maximize their net utilities given in \eqref{eq14}. It is
shown in \cite{Saraydar02} that the Nash equilibrium for this game
Pareto-dominates the SIR-balancing solution.

\subsection{Joint Power Control and Receiver Design}

The cross-layer problem of joint power control and receiver design
is studied in \cite{MeshkatiTCOM05}. It is shown that for all
linear receivers, the non-cooperative power control game in which
each user maximizes its own utility (energy efficiency) has a
unique Nash equilibrium. The equilibrium is again SIR-balanced.
The target SIR is the solution of \eqref{eq13} and is independent
of the receiver type. The results are extended to multi-antenna
systems as well. Using this non-cooperative game-theoretic
framework, the gains in energy efficiency and network capacity due
to sophisticated temporal and spatial signal processing (i.e.,
multiuser detection and multi-antenna processing) are quantified.
In particular, using a large-system analysis similar to that
presented in \cite{TseHanly99}, it can be shown that, for  the
matched filter (MF), the decorrelator (DE) and the linear MMSE
receiver, user $k$'s utility at Nash equilibrium is given by
\begin{equation}\label{eq15}
    u_k= \frac{R_k f(\gamma^*) \bar{h}_k}{\gamma^* \sigma^2}\
    \bar{\Gamma}\ ,
\end{equation}
where $\bar{\Gamma}$ depends on the receiver:
\begin{eqnarray}\label{eq16}
     &\bar{\Gamma}^{MF}& = 1-\bar{\alpha} \gamma^* \ \ \ \ \ \ \ \ \  {\textrm{for}} \ \ \bar{\alpha} < \frac{1}{\gamma*} \ , \\
\label{eq17}
    &\bar{\Gamma}^{DE}& = 1-\alpha  \ \ \ \ \ \ \ \ \ \ \ \  {\textrm{for}} \ \ \alpha < 1 \ ,  \\
\label{eq18}
    \textrm{and}  &\bar{\Gamma}^{MMSE}& = 1-\bar{\alpha} \frac{\gamma^*}{1+\gamma^*} \ \ \ {\textrm{for}} \ \ \bar{\alpha} < 1+\frac{1}{\gamma*}
    \ ,
\end{eqnarray}
with $\bar{\alpha}=\frac{\alpha}{m}$ and $\bar{h}_k = \sum_{l=1}^m
h_{kl}$. Here, $\alpha$ is the system load which is defined as the
ratio of the number of users to the processing gain (i.e., number
of users per degree of freedom), $m$ is the number of recieved
antennas and $h_{kl}$ is the channel gain from the transmit
antenna of the $k$th user to the $l$th recieve antenna. $\sigma^2$
is the noise power which includes other-cell interference.

\begin{figure}[t]
\centering \includegraphics[width=4in]{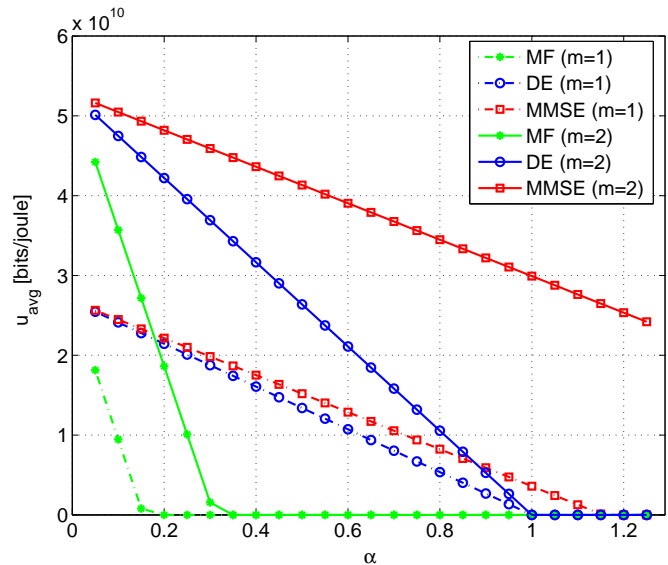} \caption{Average
utility vs. load for the matched filter (MF), the decorrelator
(DE), and the MMSE receiver with one and two receive antennas.}
\label{fig3}
\end{figure}

Fig.~\ref{fig3} shows the average utility as a function of the
system load for one and two receive antennas for a Rayleigh
channel for a user which is $100$ meters away from the uplink
receiver. The figure shows the achieved utilities for the matched
filter (MF), the decorrelator (DE) and the linear MMSE receiver.
The dashed lines correspond to single receive antenna case ($m=1$)
and the solid lines represent the case of two receive antennas
($m=2$).  It is seen from the figure that the utility (energy
efficiency) improves considerably when the matched filter is
replaced by a multiuser detector. Also, the system capacity (i.e.,
the maximum number of users that can be accommodated by the
system) is larger for the multiuser receivers as compared with the
matched filter. Among all linear receivers, the MMSE detector
achieves the highest utility. In addition, significant
improvements in user utility and system capacity are observed when
two receive antennas are used compared to the single antenna case.
The improvement is more significant for the matched filter and the
MMSE receiver as compared with the decorrelating detector. This is
because the matched filter and the MMSE receiver benefit from both
power pooling and interference reduction whereas the decorrelating
detector benefits only from power pooling (see \cite{HanlyTse01}).

\begin{figure}[t]
\centering
\includegraphics[width=4in]{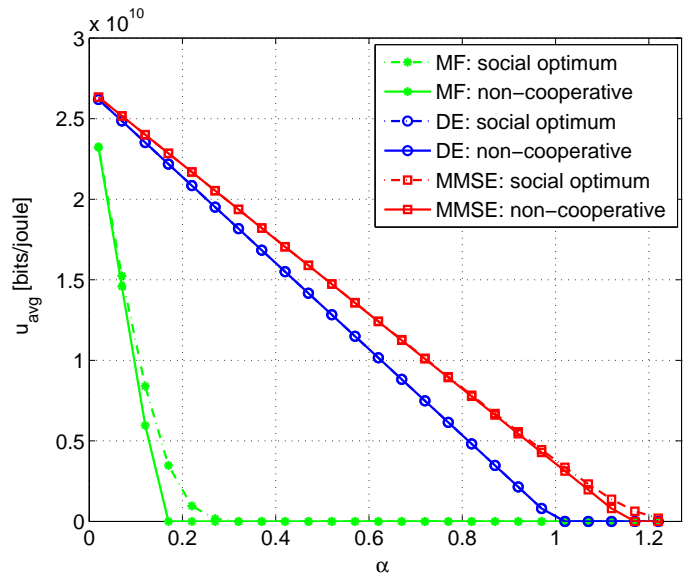}
\caption{Average utility vs load for the matched filter (MF), the
decorrelator (DE), and the MMSE receiver (single receive
antenna).} \label{fig4}
\end{figure}

Fig. \ref{fig4} shows the average utility of a user as a function
of the system load for the matched filter, decorrelator and MMSE
receivers. The solid and dashed lines correspond to the
non-cooperative and Pareto-optimal solutions, respectively. While
the difference between the non-cooperative approach and the
cooperative solution is significant for the matched filter, the
solutions are identical for the decorrelator and are quite close
to each other for the MMSE receiver. The reason is that multiuser
detectors do a better job of decoupling the users as compared to
the conventional matched filter.

\subsection{Power Control for Multicarrier CDMA}

It is well known that for maximizing the throughput, the optimal
power allocation strategy in a single-user system with parallel
AWGN channels is waterfilling \cite{CoverThomas}. The multiuser
scenario is more complicated. In \cite{Sriram01, Munz02, Shen03},
for example, several waterfilling-type approaches have been
investigated for multiuser systems to maximize the overall
throughput. However, there are many practical situations where
enhancing energy efficiency is more important than maximizing
throughput. For such applications, it is more important to
maximize the number of bits that can be transmitted per joule of
energy consumed rather than to maximize the throughput. Focusing
on a multicarrier DS-CDMA system with $D$ carriers, let us
consider a non-cooperative game in which each user chooses how
much power to transmit on each carrier to maximize its overall
energy efficiency. Let $\mathcal{G}_D=[{\mathcal{K}},
\{\mathcal{A}_k^{MC} \}, \{u_k^{MC} \}]$ denote the proposed
non-cooperative game where ${\mathcal{K}}=\{1,\cdots, K \}$, and
$\mathcal{A}_k^{MC}=[0,P_{max}]^D$ is the strategy set for the
$k${th} user. Here, $P_{max}$ is the maximum transmit power on
each carrier. Each strategy in $\mathcal{A}_k^{MC}$ can be written
as ${\mathbf{p}_k} = [p_{k1} ,\cdots, p_{kD}]$ where $p_{k\ell}$
is the transmit power of user $k$ on the $\ell$th carrier. The
utility function for user $k$ is defined as the ratio of the total
throughput to the total transmit power for the $D$ carriers, i.e.,
\begin{equation}\label{eq19}
    u_k^{MC} = \frac{\sum_{\ell=1}^D T_{k\ell}} {\sum_{\ell=1}^D p_{k\ell}} ,
\end{equation}
where $T_{k\ell}$ is the throughput achieved by user $k$ over the
$\ell${th} carrier, and is given by  $T_{k\ell} = R_k
f(\gamma_{k\ell})$ with $\gamma_{k\ell}$ denoting the received SIR
for user $k$ on carrier $\ell$. Hence, the utility-maximizing
strategy for a user is given by the solution of
\begin{equation}\label{eq20}
    \max_{{\mathbf{p}}_k} \ u_k^{MC} =  \max_{p_{k1} ,\cdots, p_{kD} } \frac{\sum_{\ell=1}^D T_{k\ell}} {\sum_{\ell=1}^D p_{k\ell}}   \ \ \ \
    \textrm{for}
   \ \  k=1,\cdots,K  ,
\end{equation}
under the constraint of non-negative powers (i.e., $p_{k\ell} \geq
0$ for all $k=1,\cdots, K$ and $\ell=1,\cdots, D$). The
multi-dimensional nature of users' strategies and
non-quasiconcavity of the utility function makes the multicarrier
problem much more challenging than the single-carrier case.

It is shown in \cite{MeshkatiJSAC06} that, for all linear
receivers and with all other users' transmit powers being fixed,
user $k$'s utility function, given by \eqref{eq19}, is maximized
when
\begin{equation}\label{eq21}
p_{k\ell}=\left\{%
\begin{array}{ll}
    p_{k L_k}^* &  \textrm{for} \ \ \ \ell=L_k\\
    0     &  \textrm{for} \ \ \ \ell \neq L_k \\
\end{array}%
\right. ,
\end{equation}
where $L_k=\arg \min_{\ell} p_{k \ell}^*$ with $p_{k \ell}^*$
being the transmit power required by user $k$ to achieve an SIR
equal to $\gamma^*$ on the $\ell${th} carrier, or $P_{max}$ if
$\gamma^*$ cannot be achieved. Here, $\gamma^*$ is the again the
solution of $f(\gamma) = \gamma \ f'(\gamma)$.

This suggests that the utility for user $k$ is maximized when the
user transmits only over its ``best" carrier such that the
achieved SIR at the output of the uplink receiver is equal to
$\gamma^*$. The ``best" carrier is the one  that requires the
least amount of transmit power to achieve $\gamma^*$ at the output
of the receiver. This solution is different from the waterfilling
solution that is obtained when maximizing simply throughput
\cite{WeiYu02}. Depending on the channel gains, the multicarrier
power control game may have no equilibrium, a unique equilibrium,
or more than one equilibrium (see \cite{MeshkatiJSAC06}).
Furthermore, with a high probability, at Nash equilibrium the
users are evenly distributed among the carriers. It is also shown
that the best-response greedy algorithm in which each user
iteratively and distributively maximizes its own utility converges
to the Nash equilibrium (when it exists).

\begin{figure}
\centering
\includegraphics[width=4in]{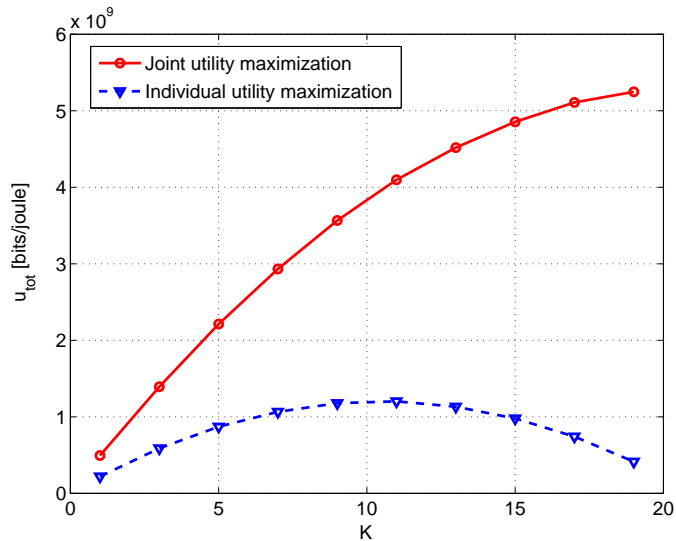}
\caption{Total utility vs. number of users, $K$, for the
two-carrier case with processing gain equal to 128.}\label{fig5}
\end{figure}

Fig.~\ref{fig5} compares the approach of joint maximization of
utility over all carriers with an approach in which the user's
utility is maximized over each carrier independently. A
significant improvement in the utility is achieved when joint
maximization over all carriers is used. This is because in the
joint optimization approach, each user transmits only on its
``best" carrier. This way, the users perform a distributed
interference avoidance mechanism which results in a higher overall
utility.

\subsection{Joint Power and Rate Control with Delay QoS Constraints}

Tradeoffs between energy efficiency and delay have recently gained
considerable attention. The tradeoffs in the single-user case are
studied in \cite{Collins99, Prabhakar01, Berry02, Fu03}. The
multiuser problem in turn is considered in \cite{Uysal02} and
\cite{Coleman04}. In \cite{Uysal02}, the authors present a
centralized scheduling scheme to transmit the arriving packets
within a specific time interval such that the total energy
consumed is minimized whereas in \cite{Coleman04}, a distributed
ALOHA-type scheme is proposed for achieving energy-delay
tradeoffs. The energy-delay tradeoff for CDMA networks is analyzed
in \cite{MeshkatiISIT05} and \cite{MeshkatiTCOMdelay} using a
game-theoretic framework.

Consider a non-cooperative game in which each user seeks to choose
its transmit power and transmission rate to maximize its energy
efficiency while satisfying its delay QoS requirements. The packet
arrival at the user's terminal is assumed to have a Poisson
distribution with an average rate of $\lambda_k$. The user
transmits the arriving packets at a rate $R_k$ (bps) and with a
transmit power equal to $p_k$ Watts. The user keeps retransmitting
a packet until the packet is received error-free. The incoming
packets are assumed to be stored in a queue and transmitted in a
first-in-first-out (FIFO) fashion. The combination of user $k$'s
queue and wireless link can be modeled as an M/G/1 queue. Now, let
$W_k$ represent the total packet delay for user $k$ including
queueing and transmission delays. We require the average delay for
user $k$'s packets to be less than or equal to $D_k$. Hence, the
proposed joint power and rate control can be expressed as the
following constrained maximization:
\begin{equation}\label{eq22}
    \max_{p_k, R_k} \ u_k \ \ \ \textrm{s.t.} \ \ \ \bar{W}_k \leq D_k \ ,
\end{equation}

It is shown in \cite{MeshkatiTCOMdelay} that the delay constraint
of a user translates into a lower bound for the user's output SIR.
Furthermore, any combination of transmit power $p_k$ and
transmission rate $R_k$ such that $\gamma_k=\gamma^*$ and $R_k\geq
\Omega_k^*$ maximizes user $k$'s utility. $\Omega_k^*$ here
corresponds to the rate at which user $k$ meets its delay
constraint with equality when $\gamma_k=\gamma^*$ and is given by
\begin{equation}\label{eq22b}
\Omega_k^*=\left(\frac{M}{D_k}\right) \frac{1+D_k\lambda_k
+\sqrt{1+ D_k^2 \lambda_k^2 +2(1-f(\gamma^*) )D_k
\lambda_k}}{2f(\gamma^*)}.
\end{equation}
This means that the joint power and rate control game has
infinitely many Nash equilibria. However, the equilibrium
corresponding to $R_k=\Omega_k^*$ with $\gamma_k=\gamma^*$ is the
Pareto-dominant equilibrium. Using this framework, the loss in
energy efficiency and network capacity due to the presence of
delay-sensitive users can be quantified. In particular, the QoS
constraints of a user can be translated into a ``size" for the
user which is an indication of the amount of resources consumed by
the user. For a matched filter receiver, at Pareto-dominant Nash
equilibrium, the ``size" of user $k$ is given by
\begin{equation}\label{eq23}
    \Phi_k^* = \frac{1}{1+\frac{B}{\Omega_k^* \gamma^*}} \ ,
\end{equation}
where $B$ is the system bandwidth. The necessary and sufficient
condition for this equilibrium to be feasible is given by
\begin{equation}\label{eq24}
    \sum_{k=1}^K \Phi_k^* < 1 .
\end{equation}
Furthermore, the utility of user $\ell$ at the Pareto-dominant
Nash equilibrium is given by (see \cite{MeshkatiTCOMdelay})
\begin{equation}\label{eq25}
u_{\ell}=\left(\frac{B h_{\ell} f(\gamma^*)}{ \sigma^2
\gamma^*}\right)\frac{1-\sum_{i=1}^K
    \Phi_i^*}{1-\Phi_{\ell}^*} \ ,
\end{equation}
where, as before, $\sigma^2$ is the noise power (including
other-cell interference) and $h_{\ell}$ is the channel gain.
Equation \eqref{eq25} together with \eqref{eq24} allows us to
quantify the tradeoffs among delay, energy efficiency, throughput
and network capacity for the multiuser, competitive setting under
consideration.

\begin{figure}
\begin{center}
\leavevmode \hbox{\epsfysize=4in \epsfxsize=4in
\epsffile{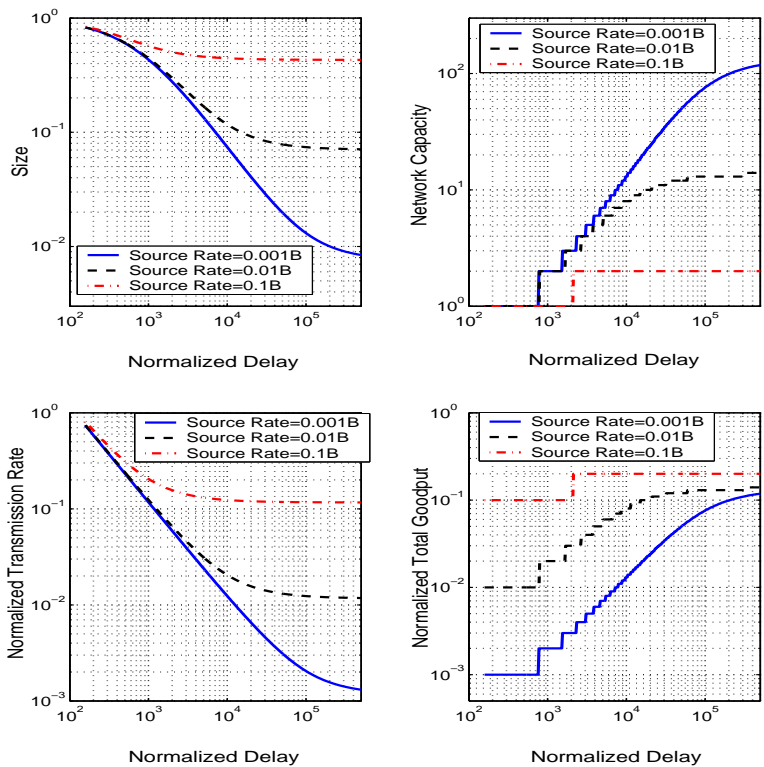}} \caption{User size, network capacity,
normalized transmission rate, and normalized total goodput as a
function of normalized delay for different source rates
($B=5$MHz).} \label{fig6}
\end{center}
\end{figure}

Fig.~\ref{fig6} shows the user size, network capacity,
transmission rate, and total goodput (i.e., reliable throughput)
as a function of normalized delay for different source
rates.\footnote{The delay is normalized by the inverse of the
system bandwidth.} The network capacity refers to the maximum
number of users that can be admitted into the network assuming
that all the users have the same QoS requirements (i.e., the same
size). The transmission rate and goodput are normalized by the
system bandwidth. The total goodput is obtained by multiplying the
source rate by the total number of users. As the QoS requirements
become more stringent (i.e., a higher source rate and/or a smaller
delay), the size of the user increases which means more network
resources are required to accommodate the user. This results in a
reduction in the network capacity. It is also observed from the
figure that when the delay constraint is loose, the total goodput
is almost independent of the source rate. This is because a lower
source rate is compensated by the fact that more users can be
admitted into the network. On the other hand, when the delay
constraint in tight, the total goodput is higher for larger source
rates.

The effect of modulation on energy efficiency has also been
studied in \cite{MeshkatiJSAC_amod} in a similar manner. In
particular, a non-cooperative game is proposed in which each user
can choose its modulation level (e.g., 16-QAM or 64-QAM) as well
as its transmit power and transmission rate. It is shown that, in
terms of energy efficiency, it is best for a user to choose the
lowest modulation level that can satisfy the user's delay QoS
constraints. This strategy is again different from the one
obtained when maximizing simply throughput. Incorporating the
choice of the modulation order into utility maximization allows us
to trade off \emph{energy efficiency} with \emph{spectral
efficiency}. For the same bandwidth and symbol rate, as a user
switches to a higher-order modulation, the spectral efficiency for
the user improves but its energy efficiency degrades (see
\cite{MeshkatiJSAC_amod} for more details).

\section{Discussions and Conclusions} \label{conclusions}

The objective of this article has been to provide an overview of
game-theoretic approaches to energy-efficient resource allocation
in wireless data networks. We have shown that game theory can be
used as a unifying framework for studying radio resource
management in wireless CDMA networks. Focusing on multiple-access
networks, we have presented a number of non-cooperative power
control games in which each user seeks to maximize its own utility
while satisfying its QoS requirements. The utility function
considered here measures the number of reliable bits transmitted
per joule of energy consumed, and is particularly useful for
energy-constrained networks. The actions open to each user in
trying to maximize its utility have been at least the choice of
transmit power and, depending on the situation, each user may also
be able to choose its transmission rate, modulation scheme, uplink
receiver type, multiantenna processing algorithm, or carrier
allocation strategy. The best-response strategies and the Nash
equilibrium solutions for these power control games have been
presented. Using this game-theoretic approach, the effects of
power control, rate control, modulation, temporal and spatial
signal processing, carrier allocation strategy and delay QoS
constraints on energy efficiency and network capacity have been
studied and quantified in a competitive multiuser setting. In
addition, it is seen that in many cases, energy-efficient resource
allocation algorithms are not spectrally efficient. Hence, there
is a clear tradeoff between maximizing energy efficiency and
maximizing spectral efficiency.

The game-theoretic framework discussed in the article is also very
suitable for studying cross-layer resource allocation in wireless
ad hoc networks and wireless local area networks (WLANs).
Non-cooperative games are very useful for analyzing ad hoc
networks due to the decentralized nature of the communication (see
\cite{Srivastava05}). Energy efficiency is also very important in
wireless ad hoc networks. However, the main challenge is to define
an appropriate utility function that captures the multihop nature
of the communication in ad hoc networks but at same time is
tractable analytically. In WLANs, users communicate to the access
point through random access schemes. A user must first compete
with other users in the network to capture the channel. Once the
channel is captured, the user will have the entire bandwidth to
itself for packet transmission. If a user is too aggressive in its
attempts for capturing the channel, it will cause many collisions
which will degrade the user's throughput. On the other hand, if
the user is too passive, it will not have access to the channel
very often and, hence, its throughput degrades. Game theory is an
effective tool for modeling the users' interactions in such a
system (see, for example, \cite{Xiao05} and \cite{LCC06}). Other
possible areas for further research are more extensive performance
comparison between non-cooperative and cooperative resource
allocation schemes, and inclusion of channel variation into the
utility maximization.


\begin{biography}{Farhad Meshkati} (S'98, M'06) received the B.A.Sc.
(Honors) degree in engineering science (electrical option) and the
M.A.Sc. degree in electrical engineering both from the University
of Toronto in Canada. He received his Ph.D. in electrical
engineering from Princeton University, NJ, in 2006. He is
currently with the Corporate R\&D Systems group at Qualcomm,~Inc.,
in San Diego, CA. His research interests are in wireless
communications, especially cross-layer radio resource management.
\end{biography}

\begin{biography}{H. Vincent Poor}  (S'72, M'77, SM'82, F'87)
is the Dean of Engineering and Applied Science at Princeton
University, where he is also the Michael Henry Strater University
Professor of Electrical Engineering. His interests lie in the area
of statistical signal processing, with applications in wireless
networks and related fields. Among his publications is the recent
book \emph{MIMO Wireless Communications} (Cambridge, 2007). Dr.
Poor is an IEEE Fellow and a member of the NAE. He is a
past-President of the IEEE Information Theory Society and past-EIC
of the \emph{IEEE Transactions on Information Theory}. Recent
recognition includes a Guggenheim Fellowship and the IEEE
Education Medal.
\end{biography}

\begin{biography}{Stuart C. Schwartz} (S'64, M'66, SM'83, F'92, LF'04)
received the B.S. and M.S. degrees from M.I.T. in 1961 and the
Ph.D.  from the University of Michigan in 1966. He is currently a
Professor of Electrical Engineering at Princeton University.  He
was chair of the department during the period 1985-1994, and
served as Associate Dean for the School of Engineering during the
period July 1977-June 1980.  During the academic year 1972-73, he
was a John S. Guggenheim Fellow and Visiting Associate Professor
at the department of Electrical Engineering, Technion, Haifa,
Israel. He has also held visiting academic appointments at
Dartmouth, University of California, Berkeley, and the Image
Sciences Laboratory, ETH, Zurich.  His principal research
interests are in statistical communication theory, signal and
image processing.
\end{biography}

\end{document}